# Ultra-high pressure disordered eight-coordinated phase of $Mg_2GeO_4$: Analogue for super-Earth mantles

Rajkrishna Dutta[1,2*], Sally J. Tracy[1], Ronald E. Cohen[1], Francesca Miozzi[1], Kai Luo[1], Jing Yang[1], Pamela C. Burnley[3], Dean Smith[4], Yue Meng[4], Stella Chariton[5], Vitali B. Prakapenka[5], and Thomas S. Duffy[2]

[1]Earth and Planets Laboratory, Carnegie Institution for Science, Washington DC 20015 USA.
[2]Department of Geosciences, Princeton University, Princeton, NJ 08544 USA.
[3]Department of Geoscience, University of Nevada, Las Vegas, NV 89154 USA.
[4]HPCAT, Advanced Photon Source, Argonne National Laboratory, Argonne, IL 60439 USA
[5]Center for Advanced Radiation Sources, University of Chicago, Chicago, IL 60637 USA.

## Abstract

$Mg_2GeO_4$ is an analogue for the ultra-high pressure behavior of $Mg_2SiO_4$, so we have investigated magnesium germanate to 275 GPa and over 2000 K using a laser-heated diamond anvil cell combined with *in situ* synchrotron X-ray diffraction and density functional theory (DFT) computations. The experimental results are consistent with a novel phase with disordered Mg and Ge, in which germanium adopts eight-fold coordination with oxygen: the cubic $Th_3P_4$-type structure. Simulations using the special quasirandom structure (SQS) method suggest partial order in the tetragonal $I\bar{4}2d$ structure, indistinguishable from $I\bar{4}3d$ $Th_3P_4$ in our experiments. These structures have not been reported before in any oxide. If applicable to silicates, the formation of this highly coordinated and intrinsically disordered phase would have important implications for the interior mineralogy of large, rocky extrasolar planets.

## Significance Statement

The work presents the first evidence for an eight coordinated phase in any silicate or its analogues using both experiments and calculations based on density functional theory. We have reported the structure, equation of state and phase stability of this novel phase. The silicate counterpart of this phase has significant implications for the structure and dynamics of interiors of large extrasolar planets.

*Corresponding author: rdutta@carnegiescience.edu

**Introduction**

A large number of exoplanets have been discovered in recent years, including many planets whose mean density indicates they have rocky interiors that may be up to ten times more massive than the Earth (1). There is significant interest in understanding the mineralogy of the deep interiors of such bodies, where the pressure at the core-mantle boundary is predicted to reach up to 1 TPa (2–4). Theoretical calculations suggest that silicate structures with partial or complete eight-fold coordination of silicon can be stabilized above 500 GPa (4). These pressures are expected to be reached within the mantles of rocky exoplanets of ~ 4 Earth masses or greater (3, 4). Phase changes with accompanying changes in cation coordination number may strongly affect the structure, dynamics, and heat flow in exoplanet interiors (1, 5).

Germanates are known to be effective analogues for silicates as they undergo similar phase transitions, but at lower pressures (6, 7). For example, the perovskite (Pv) to post-perovskite (pPv) transition occurs near 65 GPa in $MgGeO_3$ (8) as compared to ~125 GPa in the corresponding silicate. Recently, a theoretical study has examined ultra-high pressure transitions in the $MgO-GeO_2$ system and suggest that it can be an excellent analogue system for ultra-high pressure phase transitions in silicate minerals (9). These calculations predicted that $MgGeO_3$ post-perovskite and MgO will combine to form an eight-coordinated tetragonal phase of $Mg_2GeO_4$ ($I\bar{4}2d$) at ~175 GPa (9) (Suppl. Material, Fig. S1). This transition pressure is experimentally accessible with a laser-heated diamond anvil cell. The same transition is predicted to occur at ~490 GPa (4) in the silicate, which is beyond the limit of conventional static compression techniques.

Here we report laser-heated diamond anvil cell experiments to examine $Mg_2GeO_4$ above 200 GPa, demonstrating that a cubic $Th_3P_4$-type ($I\bar{4}3d$) or partially ordered $I\bar{4}2d$ phase is synthesized under such conditions. Our computations suggest an unusual isostructural transition from an almost ordered tetragonal (order parameter Q = 0.81) to almost disordered tetragonal phase (Q = 0.25) at high temperatures. Disordered ions of very different valences, $Ge^{4+}$ and $Mg^{2+}$ is highly unusual, and perhaps occurs in other high-pressure systems at high temperatures. The calculations also reveal the atomic structure and pressure-volume relationship of the disordered phase. The formation of the $Th_3P_4$-type *or* $I\bar{4}2d$ phase in the corresponding silicate at higher pressures would have important implications for the interior mineralogy of large, rocky extrasolar planets.

**Results**

In a series of experiments (Suppl. Material, Table S1), $Mg_2GeO_4$ olivine samples were compressed to pressures between 115 and 275 GPa. Prior to any heating, the X-ray diffraction (XRD) patterns contained only gold and rhenium peaks, indicating the germanate underwent pressure-induced amorphization at these conditions (10). Figure 1 shows 300-K diffraction patterns obtained on $Mg_2GeO_4$ samples that had been heated above 2000 K at pressures between 161 and 275 GPa. A consistent set of new diffraction peaks is observed in all of these patterns. The ordered tetragonal $I\bar{4}2d$ structure was predicted by DFT to be more stable than $MgGeO_3$ pPv + MgO in this pressure range (9). We carried out theoretical calculations to optimize this structure at pressures relevant to our experimental conditions. A comparison of the new peaks observed in our experiments to those of the theoretically predicted $I\bar{4}2d$-type $Mg_2GeO_4$ phase show similarities in peak positions, but the expected splittings associated with the tetragonal phase are not observed (Suppl. Material, Fig. S2). This is suggestive of formation instead of a related higher symmetry phase. Under the assumption that the unit cell of the new phase is cubic; only a single indexing is possible, consistent with a body-centered-cubic lattice ($h + k + l = 2n$).

Within the minimal supergroups of $I\bar{4}2d$ (#122), space group $I\bar{4}3d$ (#220) provides the only possible solution. Of ~40 candidate structure types for the $AB_2X_4$ stoichiometry (11), only one structure with this space group is known: the thorium phosphide structure, $Th_3P_4$ ($I\bar{4}3d$, Z=4). Fitting the observed diffraction data to this structure yields a good match to the peak positions (Suppl. Material, Table. S2).

The $Th_3P_4$-type $Mg_2GeO_4$ structure can be understood as an order-disorder transition of the tetragonal $I\bar{4}2d$ phase. The ordered tetragonal structure has two cation sites: Mg (*8d*) and Ge (*4a*), while O occupies the *16e* site. The $Th_3P_4$ structure is intrinsically disordered with both cations having partial occupancy of the *12a* Wyckoff site (Mg: 2/3, Ge: 1/3) while O occupies the *16b* site. At high temperature, a disordered phase can be stabilized relative to a corresponding ordered form by configurational entropy. While a phase exhibiting cation disorder between $Mg^{2+}$ and $Ge^{4+}$ might be considered a surprising result given the large difference in cation radius and valence, the energetics associated with different site occupancies across species may diminish under extreme compression. Firstly, we considered the relative stability of the ordered $I\bar{4}2d$ phase compared with the completely disordered cubic $I\bar{4}3d$. The disordering enthalpy increases significantly with pressure (70 – 342 GPa): $\Delta H/fu$ = 0.027-0.081 Ryd (0.37-1.10 eV). Counteracting the enthalpy of disorder is the configurational entropy, which is zero in the completely ordered tetragonal $I\bar{4}2d$ and $S/k_B/fu = 1.90954$ for the cubic disordered phase. The transition from completely ordered tetragonal to completely disordered cubic is when the Gibbs free energy difference $\Delta G = 0$, thus ranges from 2233K-6718 K for 70-342 GPa (Fig. S3).

Although many minerals and alloys have sharp first-order or fairly sharp continuous transitions from nearly ordered to disordered structures, it is also possible to have other behaviors (12). The tetragonal structure can be partly disordered, with an order parameter, *Q* that varies from 0 for the disordered $I\bar{4}3d$ structure to 1 for the completely ordered $I\bar{4}2d$ phase. The relationship is $X_{Mg}(8d) = Q \times 1/3 + 2/3$ and $X_{Mg}(4a) = -Q \times 2/3 + 2/3$. This is known as convergent ordering, because the sites become equivalent when completely disordered. On varying the order parameter from $Q = 0$ to $Q = 1$ (Suppl. Material. Section. S4), we find a phase transition. The phase transition is not the transition from a completely ordered to completely disordered structure, rather an isostructural first order phase transition within the tetragonal $I\bar{4}2d$ space group, from more ordered ($Q = 0.81$) to less ordered ($Q = 0.25$). The ΔS ordering is fairly flat (Fig. S6) from $Q = 0.2$ to $Q = 0$, whereas ΔH continues to increase, so it does not pay to completely order, even at huge temperatures (>10,000K). This comes from the large linear term in ΔH(Q) (12).

Figure 2a summarizes the transition curves for the order-disorder phase transition from this study and previous work (13). Our calculations yield transition temperatures higher than Ref. (13); insufficient information is given in that paper to determine the source of the differences, but possibly it is due to their smaller cell sizes. Increasing disorder leads to a decrease in the *c/a* ratio (Fig. 2b). The observed X-ray diffraction pattern is consistent with both the $Q = 0.25$ and 0.0 structures (Fig. 2c) and it is impossible to distinguish between the two, while the splittings expected for Q = 1 and 0.8125 are never observed. We estimate Q < ~0.53 cannot be distinguished experimentally. The diffraction patterns throughout the manuscript have therefore been indexed with the $Q = 0$ cubic $I\bar{4}3d$ Th$_3$P$_4$ structure (Fig 3, right) and has been referred to as the Th$_3$P$_4$-type phase for the experimental data. Further details of the Th$_3$P$_4$-type structure are provided in the supplementary material (Suppl. Material, Section. S3).

A heating cycle conducted at 234 GPa (H3_2) shows peaks from the Th$_3$P$_4$-type Mg$_2$GeO$_4$ phase grow with heating duration (~17 minutes heating time) and temperature (Suppl.

Material, Fig. S8). We generated a disordered special quasirandom structure (SQS) using the ATAT toolkit (14) in a 224 atom supercell and optimized the structure using Quantum Espresso (15). The average structure was obtained using FINDSYM (16). The resulting structural parameters were used as an initial model to carry out a Rietveld refinement (Figure 3, left) of the diffraction pattern quenched from peak temperature (3650 K). The corresponding two-dimensional image (Suppl. Material, Fig. S9) shows smooth diffraction rings.

The lattice parameters and atomic positions obtained from the experiments and computations are in good agreement with each other (Table. S3). The unit cell dimension from the Rietveld refinement of the experimental data at 184 GPa is 5.4930 Å, while the calculations yield 5.492 Å at 193 GPa. The Mg-O and Ge-O bond lengths from the experiments (theory) are 2.080 Å (2.015 Å) and 1.745 (1.796 Å). If the O is placed at the ideal $x$ value for the Th$_3$P$_4$ structure ($x = 1/12$), the Mg-O and Ge-O bond lengths would be 1.903 Å and 1.898 Å. Whereas the difference in the Mg-O and Ge-O bond lengths suggest a substantial distortion of the cation site, it is consistent with an increase in coordination (~5% increase in bond lengths) from the post-perovskite MgGeO$_3$ (Mg-O: 1.94 Å and Ge-O: 1.70 Å) structure (17) at 184 GPa.

The pressure-temperature conditions achieved in our experiments are shown in Figure 4. For our lowest pressure experiment (115 GPa), the diffraction peaks obtained during heating (Fig. S10) can be assigned to MgGeO$_3$ post-perovskite ($Cmcm$) and B1-MgO ($Fm\bar{3}m$), and no evidence for formation of an Mg$_2$GeO$_4$ phase was observed. These results are consistent with theoretical predictions at this pressure (Fig. S1). It is also notable that prolonged heating near 2000 K was required at this pressure to produce any diffraction peaks at all, suggesting that the decomposition of Mg$_2$GeO$_4$ into post-perovskite and periclase (MgO) may be kinetically slow due to the atomic diffusion required for the decomposition reaction.

For experiments between 130 and 170 GPa, the results are more complicated. The diffraction data are consistent with the $Th_3P_4$-type (or $I\bar{4}2d$, $Q = 0.25$) phase upon initial heating of the amorphous sample. However, upon prolonged heating and/or at higher temperatures, peaks of the post perovskite phase emerge (Fig. S11 and Table S5) and grow at the expense of $Th_3P_4$-type peaks e.g. Figure 4 (top right). In the higher-pressure runs above 175 GPa, peaks consistent with the $Th_3P_4$-type phase are observed to form and/or grow during heating in all experiments. For example, in one run (H2_2), the pPv phase was synthesized at lower pressures and then the pressure in the cell was increased to 175 GPa. Upon heating at this pressure, the post-perovskite peaks diminish over time while $Th_3P_4$-type peaks grow (Fig. 4 bottom right). Complete transformations from $Mg_2GeO_4$ to $MgGeO_3$ + MgO and vice versa are not observed because of the slow reaction kinetics. In run H4, a fresh sample was heated at 187 GPa, and only $Th_3P_4$-type peaks were observed within ~ 2 minutes of laser heating and retained even after prolonged heating to 3000 K.

These observations point to the conclusion that below 175 GPa, the new $Mg_2GeO_4$ phase is a metastable phase formed on initial heating in this pressure range. The metastable behavior can be understood as a result of transformation kinetics of the highly coordinated amorphous phase that forms on room-temperature compression. Experiments on germania glass show that the Ge-O coordination number increases with compression reaching a value greater than 7 at 92 GPa (18), and it likely continues to grow with pressure. Therefore, on initial heating at lower temperatures, it may be kinetically easier to transform to the cation-disordered, eight-fold coordinated phase of $Mg_2GeO_4$, rather than decompose into the six-fold coordinated, cation-ordered post-perovskite phase and MgO. Similar behavior has been observed in experiments on the $SiO_2$ polymorph cristobalite, in which seifertite, a phase that is thermodynamically stable at pressures above 100 GPa, was observed to form metastably on heating at pressures as low as ~11 GPa, within the stability field of stishovite (19). This was interpreted as a result of the faster kinetics for the formation of metastable seifertite from the cristobalite X-I phase. Upon prolonged heating to higher temperatures, seifertite eventually transformed into the thermodynamically stable stishovite structure, similar to the behavior we observe here.

The question remains as to why the theoretically predicted ordered $I\bar{4}2d$ phase is never observed in our experiments. The minimum temperature that could be measured on heating was ~1200 K, which combined with the expected slow transformation kinetics at such temperatures prevents us from obtaining any constraints on phase stability at low temperatures. The fast quench rate in laser-heated diamond anvil cell experiments likely precludes a transition to the tetragonal ordered phase upon cooling. Substitutional ordering requires site diffusion that may be kinetically limited at multi-Mbar pressures. In run H4, we attempted to decrease the laser power in small steps to replicate slow cooling (3032 K to 300 K in ~14 minutes). The ordered $I\bar{4}2d$ phase was still not observed. We should also consider the instrumental, thermal, and possible site disorder broadening, which may make it difficult to distinguish the tetragonal phase splittings. Further work will be necessary to better determine if the $I\bar{4}2d$ phase can be synthesized at lower temperatures.

Figure 5a shows the pressure-volume relation of the $Th_3P_4$-type $Mg_2GeO_4$ phase at 300 K (0 K theoretically). No pressure-transmitting medium was used in these experiments to maximize the sample volume at extreme pressures. To minimize the effect of differential stresses on the experimental results, we have included only data points obtained immediately after quenching from high-temperature at each pressure step. Theoretical calculations were also performed to test the reliability of the equation of state parameters. A third-order Birch-Murnaghan fit (solid:

experiments; dashed: theory) to the data yields $V_0 = 252.5$ (3.2) Å$^3$, $K_{0T} = 188$ (11) GPa and $V_0 = 261.6$ (1.2) Å$^3$, $K_{0T} = 170$ (3) GPa respectively; $K'_{0T}$ was fixed at 4 in both cases. The ambient-pressure bulk modulus is comparable to that of the pPv (8) MgGeO$_3$ phase ($K_{0T} = 192$ (5) GPa, $K'_{0T} = 4$).

Figure 5b compares the densities of phases in the MgO-GeO$_2$ system (17, 20–25). The densities of MgGeO$_3$ + MgO mixtures were calculated assuming an ideal solid solution (i.e., the molar volumes are additive). The density of the mixture ($\rho$) is given by:

$$\frac{1}{\rho(P,T)} = \frac{(1-m_2)}{\rho_1(P,T)} + \frac{m_2}{\rho_2(P,T)},$$

where $m$ and $\rho$ are the mass fraction and density of the individual components. Our results indicate that the transition from post-perovskite and MgO to the Th$_3$P$_4$-type phase produces a substantial (2.4%) change in density ($\Delta\rho$) at 190 GPa.

**Discussion**

Changes in silicon coordination in minerals can influence the structure and dynamics of planetary interiors. The transition in silicates from tetrahedral (four-fold) to octahedral (six-fold) coordination of silicon by oxygen occurring near 660-km depth defines the major structural boundary in Earth's mantle (26). In the Earth, six-coordinated silicate phases are expected to be stable throughout the lower mantle with (Mg, Fe)SiO$_3$ post-perovskite and SiO$_2$ seifertite ($\alpha$-PbO$_2$-type, $Pbcn$) stable at core-mantle boundary conditions (~135 GPa). In the case of SiO$_2$, a further transition to the pyrite-type phase ($Pa\bar{3}$) is observed at 268 GPa (27), but this phase retains 6-fold coordination (sometimes described as 6 + 2). Increases in coordination above six have been reported in SiO$_2$ glass at ultra-high pressures (28). However, there is no experimental evidence for silicon coordination greater than six in any high-pressure crystalline silicates or their analogues.

While pPv MgSiO$_3$ is the expected silicate phase at the base of Earth's mantle, post-pPv phases with Si-O coordination > 6 may play a decisive role in terrestrial super-Earths of four or greater Earth masses (Suppl. Material, Fig. S1). If the Th$_3$P$_4$-type or the closely related partially disordered $I\bar{4}2d$ phase also exists in Mg$_2$SiO$_4$, the change in coordination from 6-fold to 8-fold may be accompanied by major changes in physical, chemical, and thermodynamic properties. The ~2.4% volume change associated with the transition is large ($\Delta\rho$ = ~1.5% for Pv to pPv) for ultra-high pressure phase transitions and may affect the dynamic behavior of the deep mantle.

The Th$_3$P$_4$-type phase is related to the proposed tetragonal $I\bar{4}2d$ structure by an order-disorder transition across cation sites. The expected high internal temperatures of super-Earth mantles (Suppl. Material, Fig. S1) will likely favor the Th$_3$P$_4$-type phase or mostly disordered $I\bar{4}2d$ structure ($Q = 0.25$) over the completely ordered $I\bar{4}2d$ structure ($Q = 1$). Its intrinsically disordered nature may also be an important feature as it suggests enhanced miscibility of chemical components at ultrahigh P-T conditions. Among chalcogenides, Th$_3$P$_4$-type phases are noted for their flexible structure and prevalence of defects, impurities, and disorder (29). This structural flexibility can affect rheology and transport properties. Compared to other structures, enhanced phonon scattering in this disordered phase may yield an anomalously low thermal conductivity (30).

Based on the calculated 0-K transition pressure from pPv to the ordered $I\bar{4}2d$ phase (4), it is expected that a transition from six-fold coordinated silicate phases to eight-fold coordinated phases would occur in terrestrial exoplanets greater than about four Earth masses in size. Thus, the interior dynamics of planets above and below this size range may show important differences due to the structural phase transition.

In summary, $Mg_2GeO_4$ adopts the thorium phosphide structure or a highly disordered (Q = 0.25) tetragonal $I\bar{4}2d$ phase at pressures above ~190 GPa. The phase is calculated to be ~ 2.4% denser than the mixture of $MgGeO_3$ post-perovskite and MgO. This is the first experimental synthesis of a phase with Ge in eight-fold coordination with oxygen and the first occurrence of the $Th_3P_4$-type phase in an oxide. Just as the discovery of widespread six-coordinated germanates/silicates profoundly altered our understanding of silicate crystal chemistry and its role in the Earth's deep interior, the discovery of an eight-fold coordinated, intrinsically disordered germanate opens the possibility of previously unexplored crystal-chemical behavior in the silicate minerals of large rocky exoplanets. Our results also raise the possibility that this structure or the related $I\bar{4}2d$ structure could also be adopted by other oxide minerals at extreme pressures. Thus, our results suggest the possibility of novel crystal chemistry in $A_3O_4$- and $AB_2O_4$-type compounds that warrant further exploration.

**Acknowledgments**

We acknowledge helpful discussions with Mike Walter. We are also grateful to Peng Ni for his assistance with sample synthesis. This research was supported by the National Science Foundation (NSF) – Earth Sciences (EAR-1836852). RD is grateful for support from the Carnegie Endowment. We acknowledge the support of GeoSoilEnviroCARS (Sector 13), which is supported by the NSF - Earth Sciences (EAR-1128799), and the U.S. Department of Energy (DOE), Geosciences (DE-FG02-94ER14466). Portions of this work were performed at HPCAT (Sector 16), Advanced Photon Source (APS), Argonne National Laboratory. HPCAT operations are supported by DOE-NNSA's Office of Experimental Sciences. This research used resources of the Advanced Photon Source, a DOE Office of Science User Facility operated by Argonne National Laboratory under Contract No. DE-AC02-06CH11357. REC gratefully acknowledges the Gauss Centre for Supercomputing e.V. (www.gausscentre.eu) for funding this project by providing computing time on the GCS Supercomputer SuperMUC-NG at Leibniz Supercomputing Centre (LRZ, www.lrz.de).

**Experimental methodology**

Mg$_2$GeO$_4$ olivine samples were synthesized according to established procedures (21, 31) and confirmed by X-ray diffraction. The sample was mixed with 10 wt.% gold which acted as a pressure calibrant and laser absorber. Rhenium gaskets were pre-indented and ~20 μm diameter holes were laser drilled to form the sample chamber in a diamond anvil cell. The sample + Au pellets were then loaded into a cell with beveled 50-μm culet anvils. No pressure medium was used in order to maximize the X-ray diffraction signal from the sample.

Angle-dispersive X-ray diffraction was carried out at 13-ID-D and 16-ID-B of the Advanced Photon Source using monochromatic X-rays ($\lambda$ = 0.3344 Å and 0.4066 Å, respectively) focused to dimensions of 3 x 3 μm$^2$ and 5 x 3 μm$^2$, respectively. CdTe 1M Pilatus detectors were used to collect the diffraction patterns. LaB$_6$ and CeO$_2$ standards were used to calibrate the detector position and orientation.

High temperatures were attained by double-sided heating with diode pumped fiber lasers with ~10-15 μm spot sizes. Temperature was increased in a stepwise fashion with ~150-200 K steps with a heating duration of 1.5-2 minutes at each step (total heating duration: 7-27 minutes). The laser power was adjusted independently on both sides to keep temperature differences to < 150 K. Temperatures were measured using spectroradiometry. Pressure was determined using the (111) reflection of Au (32). Thermal pressures were accounted for using the Mie-Grüneisen equation of state.

The two-dimensional X-ray diffraction images were integrated to one-dimensional patterns using DIOPTAS (33) and fit using background-subtracted Voigt line shapes. Lattice parameters were calculated by least-squares refinement of the peak positions (34) or whole profile Le Bail refinement as implemented in MAUD (35). For the whole pattern refinement, the background was fit with a 4th order polynomial, instrumental profile terms were fit with Gaussian Cagliotti terms and sample broadening incorporated isotropic size and strain broadening.

**Computational Details**

We performed first-principles density functional theory calculations to determine the structural parameters of the $I\bar{4}2d$ and Th$_3$P$_4$-type Mg$_2$GeO$_4$ phases, constrain the equation of state over the range from 135 - 342 GPa and study the details of the order-disorder transition between the two phases. We simulated the disordered system using the special quasirandom structure method (36, 37) as implemented in the ATAT-toolkit (14). We generated a 224-atom supercell and used the Monte Carlo method to find the most random structure considering clusters up to 6 neighbors as a distance of up to 3.2 Å for a cubic lattice. We constrained the supercell to be doubled in each direction of the cubic cell, i.e. 10.984 Å × 10.984 Å × 10.984 Å for our starting system. QUANTUM ESPRESSO (15) plane wave pseudopotential code PWSCF was used to perform DFT computations, with PBE exchange-correlation, and PAW potentials from the pslibrary pseudopotential library (38) with a 4x4x4 Monkhorst-Pack k-point mesh in the supercell and an energy cutoff for the wavefunctions of 100 Ryd.

**Fig. 1** X-ray diffraction patterns of the Th$_3$P$_4$-type phase of Mg$_2$GeO$_4$ at 161 GPa, 225 GPa, and 275 GPa. The two lower pressure patterns are collected upon quench to 300 K after heating to 1806 K and 3650 K respectively, whereas the upper pattern was collected *in situ* at 2020 K. Peaks from Th$_3$P$_4$-type Mg$_2$GeO$_4$ are indicated with Miller indices.

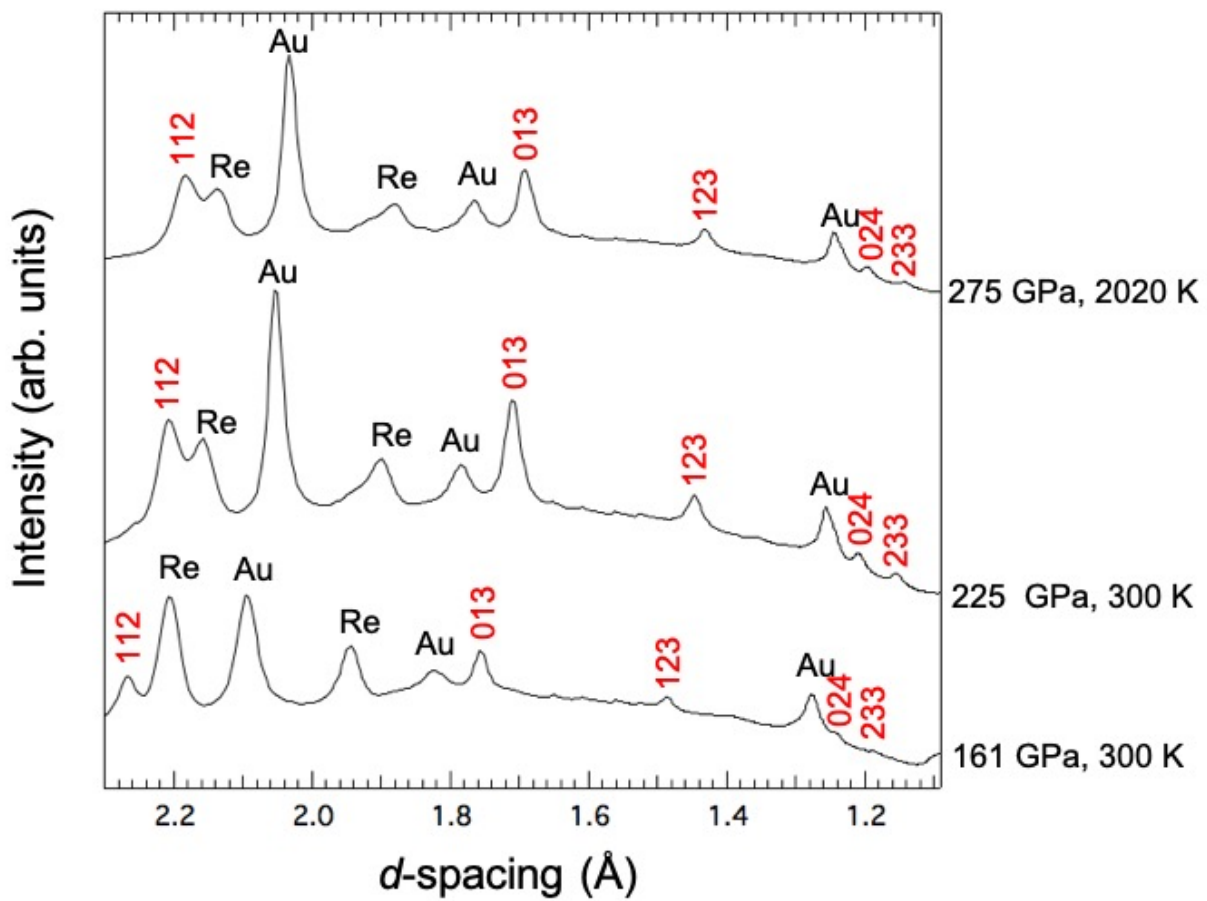

**Fig. 2(a)** Order-disorder transition between $I\bar{4}2d$ and $I\bar{4}3d$ $Mg_2GeO_4$ from DFT calculations (solid black line). The red circles indicate the isostructural transition from mostly ordered (Q = 0.8125) to mostly disordered (Q = 0.25) $I\bar{4}2d$ structure. The dashed line shows the transition curve from Ref. 16. **(b)** Variation in $c/a$ with order parameter at 200 GPa; triplets, quadruplets and sextuplets are the different SQS cluster sizes. **(c)** Comparison of the observed X-ray diffraction pattern (black) at 187 GPa, 2010 K with the theoretically calculated structures with Q = 0.8125 (blue), Q = 0.25 (green) and Q = 0 (red) at 200 GPa. The FWHM of the observed sample peaks has be used as Gaussian broadening for the simulated patterns.

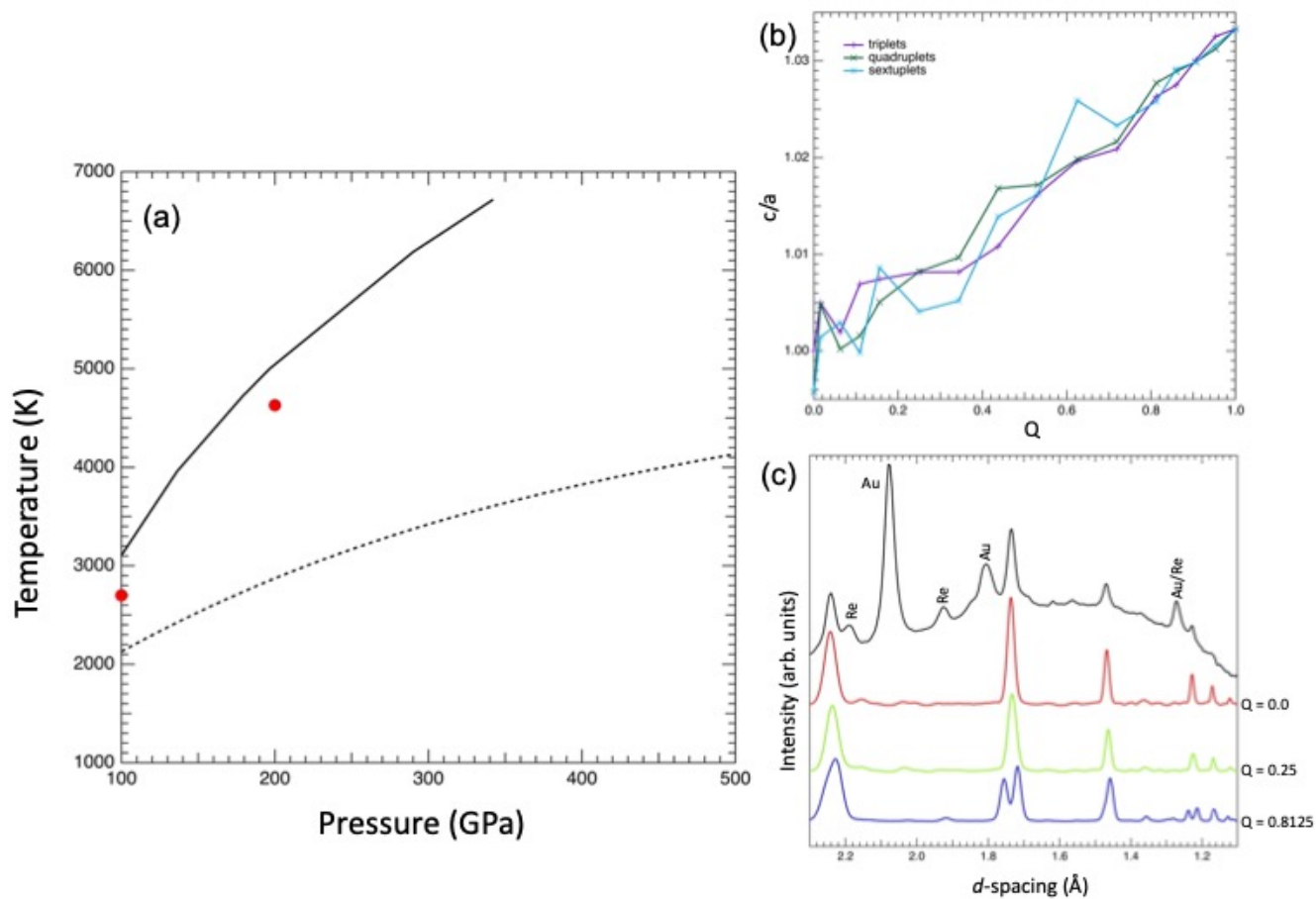

**Fig. 3(a)** Rietveld refinement of the X-ray diffraction pattern (black) obtained after heating $Mg_2GeO_4$ to 3650 K at 240 GPa with subsequent quenching to room temperature ($a$ = 5.4055 (2) Å, $O_x$ = 0.0468). Red and green curves show the fit and background, respectively. Miller indices of the $Th_3P_4$-type $Mg_2GeO_4$ are shown above the observed diffraction pattern. A small amount of untransformed starting material is present near 1.95 A. **(b)** Structure of the $Th_3P_4$-type phase: Red spheres are oxygen atoms and orange/mauve spheres represent the Mg/Ge site with shading to indicate 2/3 occupancy by Mg and 1/3 occupancy by Ge. A representative $(Mg,Ge)O_8$ polyhedron is also shown.

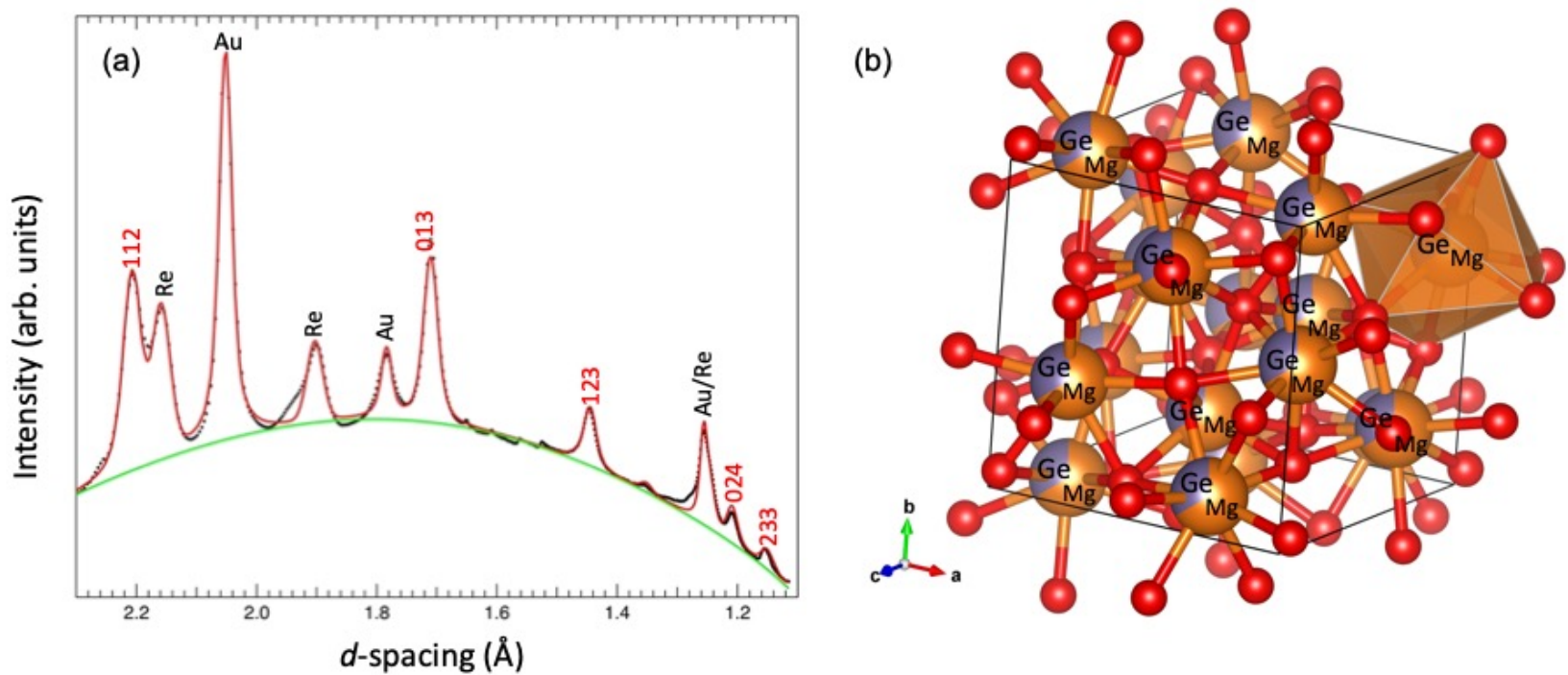

**Fig. 4(a)** Phase diagram of $Mg_2GeO_4$. Solid brown circles indicate P-T conditions where only $MgGeO_3$ pPv + MgO is observed. The solid red circles show P-T conditions where diffraction patterns showed only $Th_3P_4$-type $Mg_2GeO_4$ phase. The open red circles show lower temperature conditions at which $Th_3P_4$ phase is observed, but is interpreted as metastable. The solid grey symbols indicate conditions at which diffraction from both pPv and $Th_3P_4$ are observed, but the pPv peaks grow and the $Th_3P_4$ peaks diminish with increasing heating time and/or temperature (panel b). The growth of pPv at the expense of $Th_3P_4$ is taken as evidence that pPv is the stable phase at these conditions. Conversely, the solid pink symbols show conditions at which the peaks of the $Th_3P_4$ phase are observed to grow and the pPv peaks diminish with heating time and/or temperature (panel c). This is evidence that the $Th_3P_4$ phase is stable at these conditions. The black line is the estimated phase boundary for the pPv + MgO→ $I\bar{4}3d$ $Mg_2GeO_4$ from this work. Open brown diamonds are P-T points at which pPv was observed in previous work (17). The stability regions for pPv+MgO, pPv + MgO with metastable $Th_3P_4$ at lower temperatures, and $Th_3P_4$ are indicated at the bottom. **(b)** Diffraction patterns during a heating cycle at 169 GPa showing the growth of post perovskite peaks (brown circles) at the expense of $Th_3P_4$ peaks (red circles). **(c)** Diffraction patterns during a heating cycle at 175 GPa showing the growth of $Th_3P_4$-type phase (indicated by red circles) and decreased intensity of the pPv phase (indicated by brown circles).

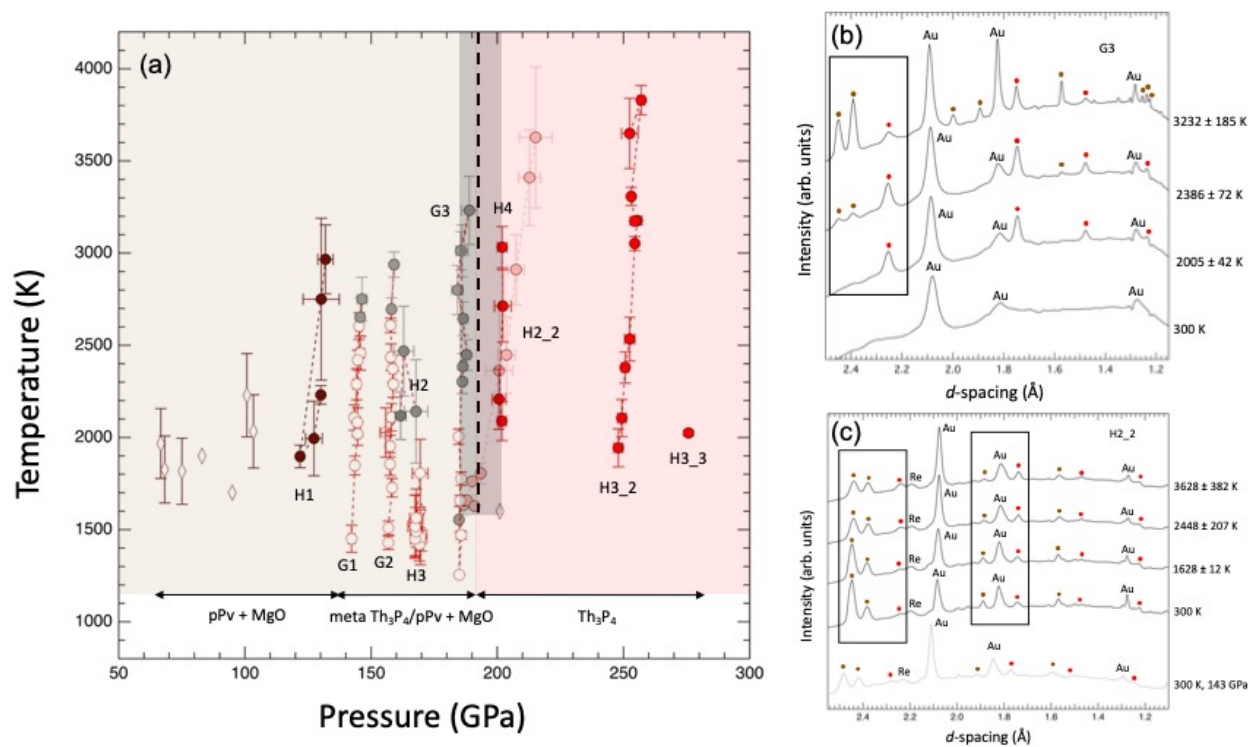

**Fig. 5(a)** Equation of state of the Th$_3$P$_4$-type phase from experiments (300 K, solid red circles) and theoretical calculations (0 K, open red circles). Lines are 3$^{rd}$ order Birch-Murnaghan equation fits. **(b)** Densities of selected phases in the MgO-GeO$_2$ system. Solid circles are from this work (brown: pPv; red: Th$_3$P$_4$). Previous studies are indicated by open symbols, brown diamonds: pPv (16); green triangles: Pv (20); purple square: spinel (21, 22); black stars: olivine (23); blue hexagons: B1-MgO (24, 25). The brown and green solid lines are the calculated densities of mixtures of MgGeO$_3$ Pv and pPv with MgO.

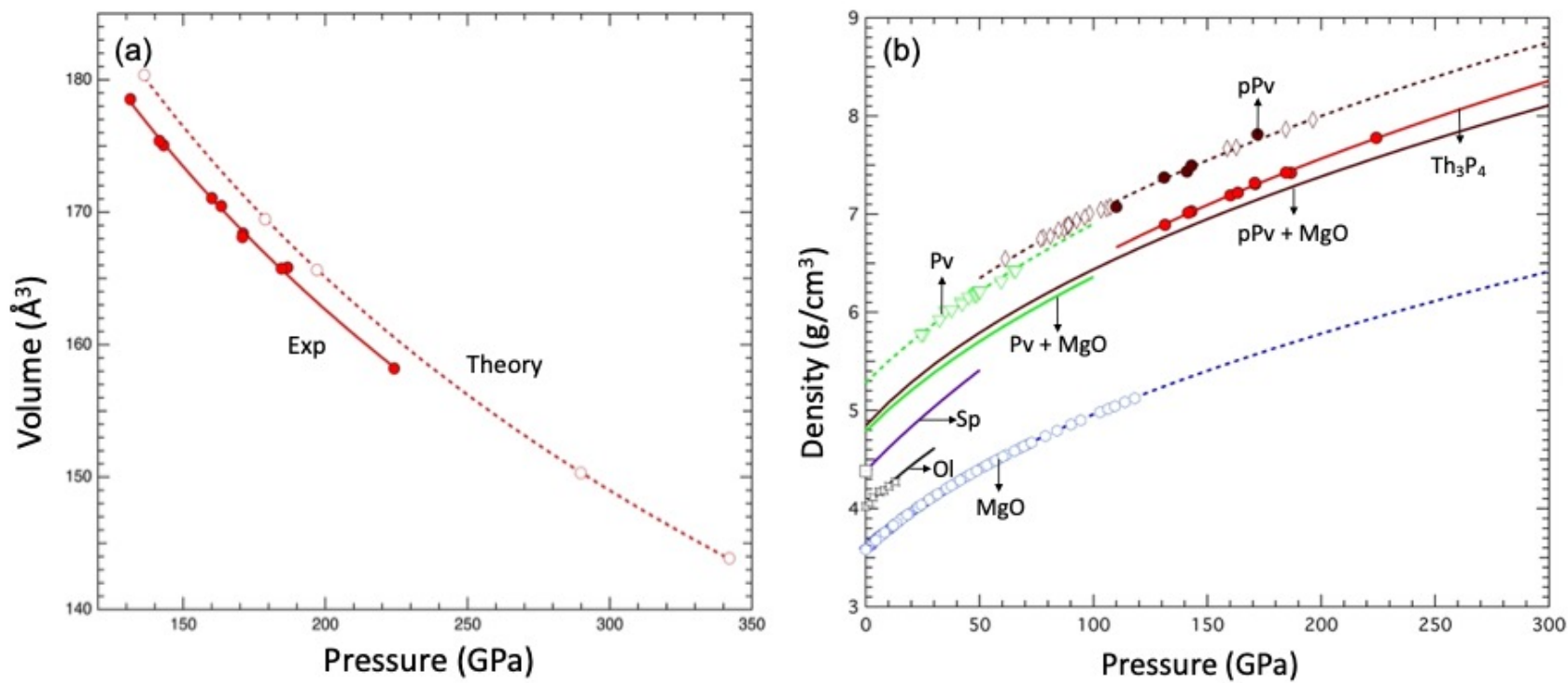

# Supplementary Material

**Ultra-high pressure disordered eight-coordinated phase of $Mg_2GeO_4$: Analogue for super-Earth mantles**

Rajkrishna Dutta[1, 2*], Sally J. Tracy[1], Ronald E. Cohen[1], Francesca Miozzi[1], Kai Luo[1], Jing Yang[1], Pamela C. Burnley[3], Dean Smith[4], Yue Meng[4], Stella Chariton[5], Vitali B. Prakapenka[5], and Thomas S. Duffy[2]

[1]Earth and Planets Laboratory, Carnegie Institution for Science, Washington DC 20015 USA.
[2]Department of Geosciences, Princeton University, Princeton, NJ 08544 USA.
[3]Department of Geoscience, University of Nevada, Las Vegas, NV 89154 USA.
[4]HPCAT, Advanced Photon Source, Argonne National Laboratory, Argonne, IL 60439 USA
[5]Center for Advanced Radiation Sources, University of Chicago, Chicago, IL 60637 USA.

**Contents**

## S1. Predicted phase boundaries for $Mg_2GeO_4$ and $Mg_2SiO_4$ at multi-megabar pressures

**Fig. S1.** Theoretically calculated phase boundaries (1, 2) for $Mg_2SiO_4$ (grey solid line) and $Mg_2GeO_4$ (black solid line). Orange and blue squares show the pressure-temperature conditions expected at the core-mantle boundaries of terrestrial- and water-rich exoplanets, respectively (3). The numbers in each square represent the mass of the planet as a multiple of Earth's mass. The shaded region represents the P, T range of this study. The question mark indicates the possible stability of $Th_3P_4$-type $Mg_2SiO_4$. Modified after Umemoto et al., 2017, 2020.

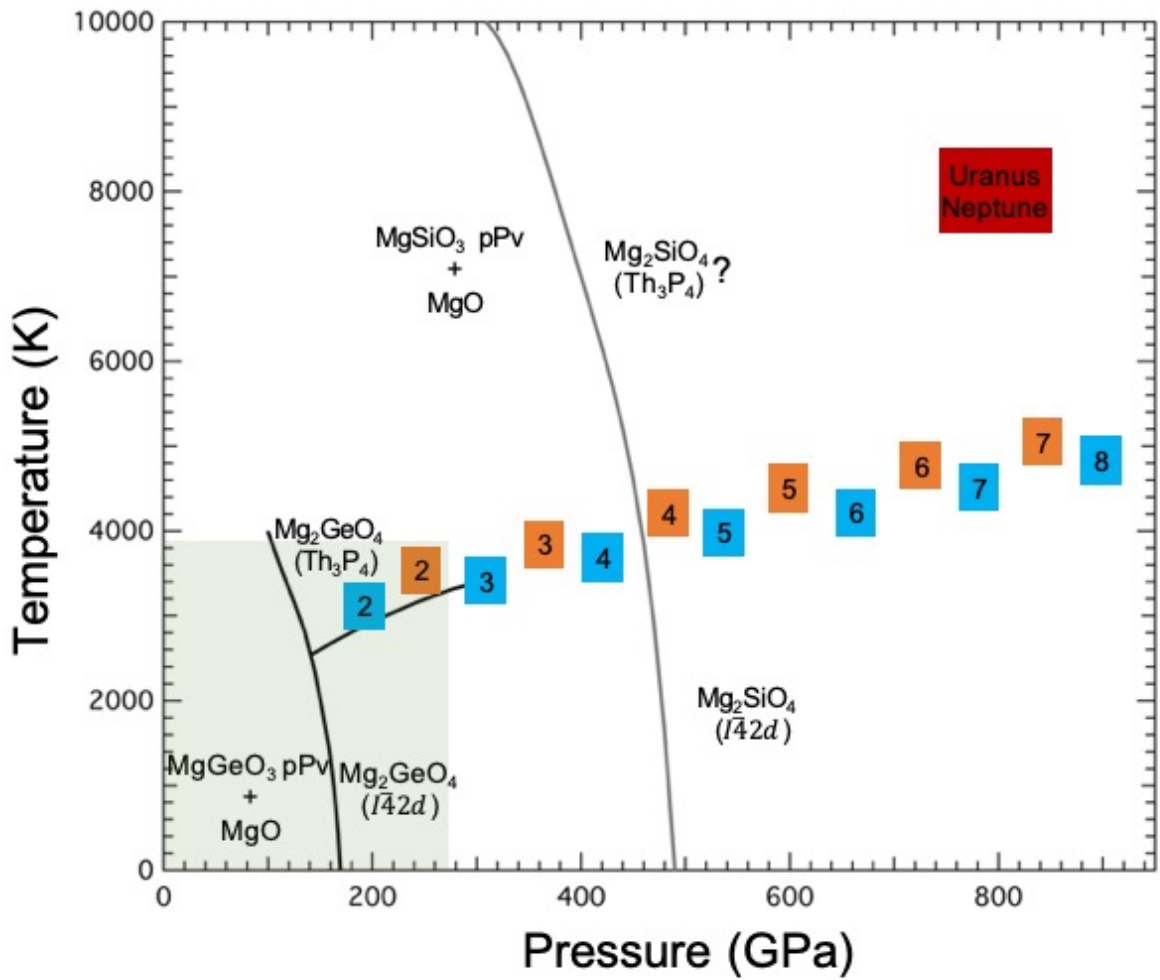

## S2. Summary of Experimental runs

**Table S1.** Summary of the experimental conditions and results

| Run. No | Pressure (GPa) | Peak Temperature (K) | Heating duration (mm: ss) | Phase(s) Before Heating | Phase(s) Observed During Heating |
|---|---|---|---|---|---|
| H1 | 115 | 2966 ± 186 | 26: 42 | Amorphous | pPv |
| G1 | 134 | 2751 ± 118 | 25:47 | Amorphous | $Th_3P_4$ (↓) + pPv (↑) |
| G2 | 151 | 2938 ± 69 | 13:09 | Amorphous | $Th_3P_4$ (↓) + pPv (↑) |
| H2 | 153 | 2468 ± 243 | 10:05 | Amorphous | $Th_3P_4$ (↓) + pPv (↑) |
| H3 | 160 | 1806 ± 184 | 13:47 | Amorphous | $Th_3P_4$ |
| G3 | 169 | 3232 ± 185 | 11:12 | Amorphous | $Th_3P_4$ (↓) + pPv (↑) |
| H2_2 | 175 | 3628 ± 382 | 21:12 | $Th_3P_4$ + pPv | pPv (↓)+ $Th_3P_4$ (↑) |
| H4 | 187 | 3032 ± 113 | 13:32 | Amorphous | $Th_3P_4$ |
| H3_2 | 234 | 3650 ± 191 | 17:05 | $Th_3P_4$ | $Th_3P_4$ |
| H3_3 | 261 | 2025 ± 45 | 07:21 | $Th_3P_4$ | $Th_3P_4$ |

pPv – $MgGeO_3$ post-perovskite phase; $Th_3P_4$ – thorium phosphide-type $Mg_2GeO_4$ phase. Upward and downward pointing arrows indicate that the diffraction peaks of the given phase grew or diminished, respectively, as the heating temperature and/or time was increased. The growth of one phase and reduction of another is used to identify the stable phase as low reaction rates under these conditions prevent complete transformation on the timescale of the heating experiments.

## S3. Structure and crystal chemistry of the thorium phosphide-type phase

The $Th_3P_4$ structure was originally described by Meisel (1939) (4) and Kripyakevich (1963) (5). It is based on a body-centered-cubic lattice with the cations in eight-fold coordination. The Mg and Ge ions occupy site *12a*, while the oxygens are located on site *16c*. Thus, the structure is intrinsically disordered with both cations occupying the same crystallographic site. There is a single positional parameter, $x$. The nearest-neighbor cation-anion bond distances are identical when $x$ adopts its ideal value, 1/12. The cation polyhedra are in the form of a strongly distorted cube (octaverticon) with triangular faces. Each anion is bonded to six cations in a distorted octahedron.

The $Th_3P_4$-type structure occurs commonly in binary chalcogenides of the light rare earth elements in both defective ($A_2X_3$) and non-defective forms ($A_3X_4$), often exhibiting solid solutions between them (6). It also occurs widely in ternary $AB_2X_4$ chalcogenides where A is a divalent metal cation and B is a light rare earth element (7, 8). Its prevalence in light rare earths indicate the structure is favored by compounds with a large ratio of cation to anion radius. Based on $AB_2X_4$ structure maps (7, 8), the phase is restricted to compounds in which the radii of the cations, $r_A$ and $r_B$, are similar ($r_A/r_B \sim 0.8 - 1.2$) and have low electronegativity. At larger cation radius ratios, the calcium ferrite-type structure ($CaFe_2O_4$) is preferred instead. Chalcogenides with the $Th_3P_4$ structure have attracted interest for technical applications because the structure is highly flexible and can incorporate many types of impurities and defects (9).
At high pressures and temperatures, ternary sulfospinels $AB_2S_4$ (A=Mg, Mn, B=Tm, Yb) have been shown to transform to the $Th_3P_4$-type structure (10). The structure has also been synthesized at high pressure in nitrides including $Hf_3N_4$, $Zr_3N_4$, and $Ti_3N_4$, in which the phase is found to be highly incompressible (11, 12). The structure can also describe an electride phase in alkali metals formed at high pressure (13).

**Fig. S2.** Comparison of the observed diffraction pattern (black) at 187 GPa, 2010 K with the theoretically calculated $I\bar{4}2d$- (green ticks; $a$ = 5.417 Å and $c$ = 5.592 Å at 197 GPa) and $I\bar{4}3d$-type $Mg_2GeO_4$ structures (red ticks; $a$ = 5.492 Å at 193 GPa). Miller indices of $Th_3P_4$-type (red) $Mg_2GeO_4$ are listed above each peak. Red and green curves show the fit and background, respectively.

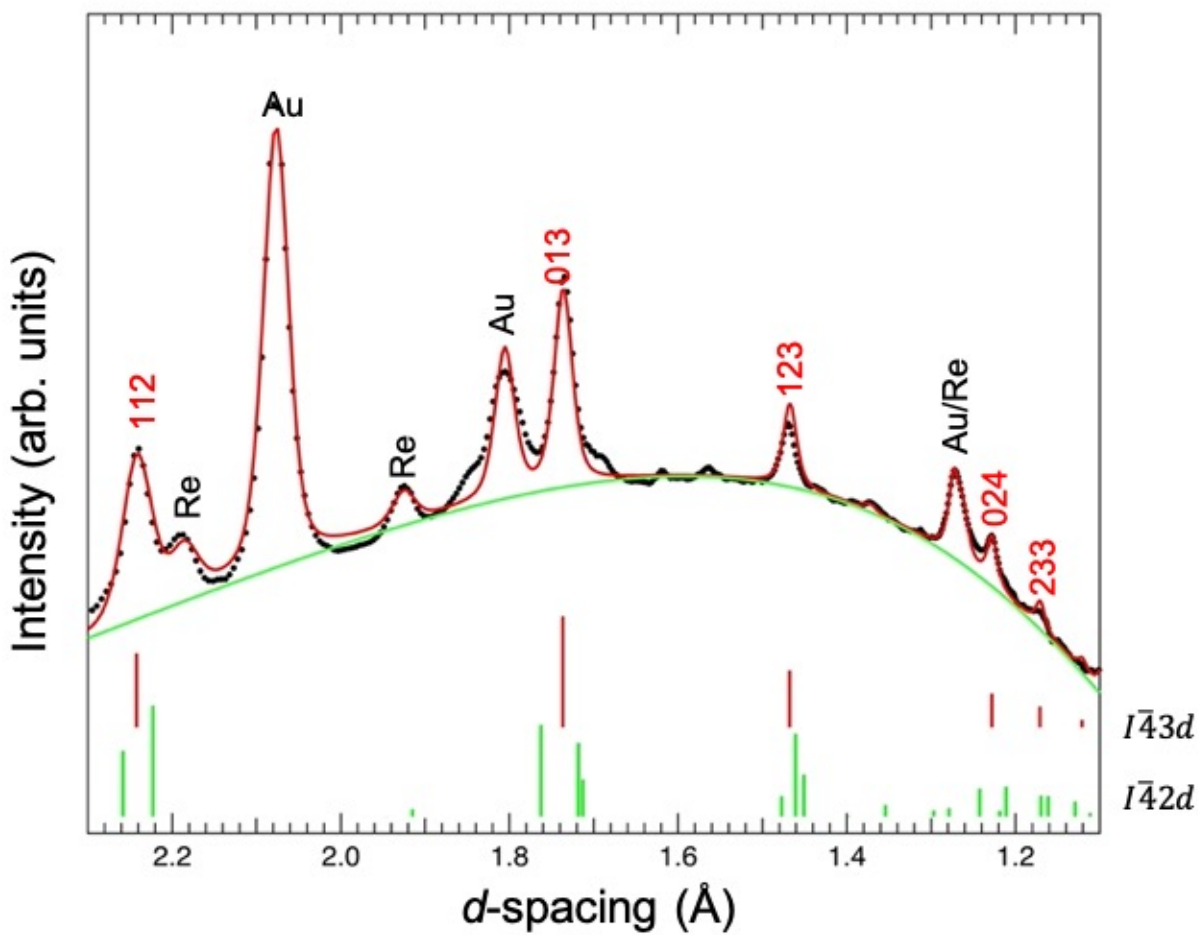

**Table S2.** Calculated and observed *d*-spacings of Th$_3$P$_4$-type Mg$_2$GeO$_4$ obtained after heating to 3650 K at 240 GPa and then quenching to room temperature. The lattice parameter is *a* = 5.4055 (7) Å. The unit cell dimension is in excellent agreement with that obtained from a Rietveld refinement of the same pattern, *a* = 5.4055 (2) Å.

| *h* | *k* | *l* | *d*-observed (Å) | *d*-calculated (Å) | Δ*d* (Å) |
|---|---|---|---|---|---|
| 1 | 1 | 2 | 2.2064 | 2.2068 | -0.0004 |
| 0 | 1 | 3 | 1.7090 | 1.7093 | -0.0003 |
| 1 | 2 | 3 | 1.4462 | 1.4446 | 0.0016 |
| 0 | 2 | 4 | 1.2079 | 1.2087 | -0.0008 |
| 2 | 3 | 3 | 1.1525 | 1.1524 | 0.0001 |

**Table S3.** Structural parameters of the Th$_3$P$_4$-type Mg$_2$GeO$_4$ from experiments and DFT calculations.

|  | Experiment | Theory |
|---|---|---|
| Pressure (GPa) | 184 | 193 |
| *a* (Å) | 5.4930 (7) | 5.492 |
| Mg/Ge | (0.375, 0, 0.25) | (0.375, 0, 0.25) |
| O | (0.055, 0.055, 0.055) | (0.0646, 0.0646, 0.0646) |
| Mg-O (Å) | 1.745 | 1.796 |
| Ge-O (Å) | 2.080 | 2.015 |

## S4. Order-disorder transition

*Static DFT calculations*

Within DFT, we relaxed the ordered structures with respect to lattice parameters and atomic positions by optimizing the enthalpy at the same pressures that were determined for the special quasirandom structures (SQS) models of the cubic structure, giving seven points with pressures ranging from 70-342 GPa. Fig. S3 shows the difference in Gibbs free energy between the ordered and disordered structures at different temperatures.

*Order Parameter*

We generated a set of SQS structures for $I\bar{4}2d$-type Mg$_2$GeO$_4$ for 16 different order parameters from $Q = 0$ to 1 for 224 atom 2x2x2 supercells (e.g. Fig. S4). In this case, we performed SQS for clusters up to 3rd, 4th, and 6th neighbors. We used the Γ-point and performed static relaxations at pressures of 200 and 100 GPa for each of the 16 $Q$'s. The enthalpies with different cluster sizes do not vary systematically with cluster size (Fig. S5) and thus should instead be considered as 3 different samples of possible random structures. We can simply fit the resulting enthalpies versus Q to estimate the state of order versus T at each pressure. The entropy is given by $\frac{S}{k_B} = \sum X \ln X$, where the sum is over the sites and X is the mole fraction of Mg or Ge on each site (Fig. S6). The free energy is minimized at each temperature, giving the order parameter versus temperature (Fig. S7). The order of the transition is strongly dependent on the order of fit to the enthalpy versus Q. A second order fit yields gradual disordering with no transition, but 3$^{rd}$ or higher order fits (which better fit the data, Table S4.) show a sharp first order isostructural transition.

*First principles molecular dynamics*

We can also obtain estimates of free energies from MD simulations, but these are very computationally intensive (about 500 times more computer time is needed), on what are already large computations. Thus only a few points have been done—the fully disordered cubic SQS structure, the fully ordered $I\bar{4}2d$ phase, both at the same static pressure of 196 GPa, a = 5.492 Å for cubic and $a$ = 5.416 Å $c$ = 5.592 Å for tetragonal (the MD supercells are double these in each direction). Here we used the GBRV pseudopotentials (14) with an ecut of 40 Ryd and $\Gamma$-point only for efficiency. We performed NBVT MD with the svr stochastic thermostat and a time step about 1 fs. The run was 5 ps with the first ps removed for equilibration. The MD pressures for cubic and tetragonal at 5000 K are 240.2 and 241.8 GPa, respectively, showing that the thermal pressure does not depend strongly on the state of order. We also attempted the same simulation for $Q$ = 0.81 and $Q$ = 0.25, and no clear trends in vibrational entropy with $Q$ were observed. More configurations and larger cell sizes are needed to compute the phase diagram, but it is clear from the static computations (and our experiments) that $Mg_2GeO_4$ disorders at high temperatures at these pressures.

**Fig. S3** Difference in Gibbs free energy between completely ordered $I\bar{4}2d$- and completely disordered $I\bar{4}3d$- $Mg_2GeO_4$ at different temperatures. The dashed grey line indicates $\Delta G = 0$ shows the transition temperatures at the concerned pressure.

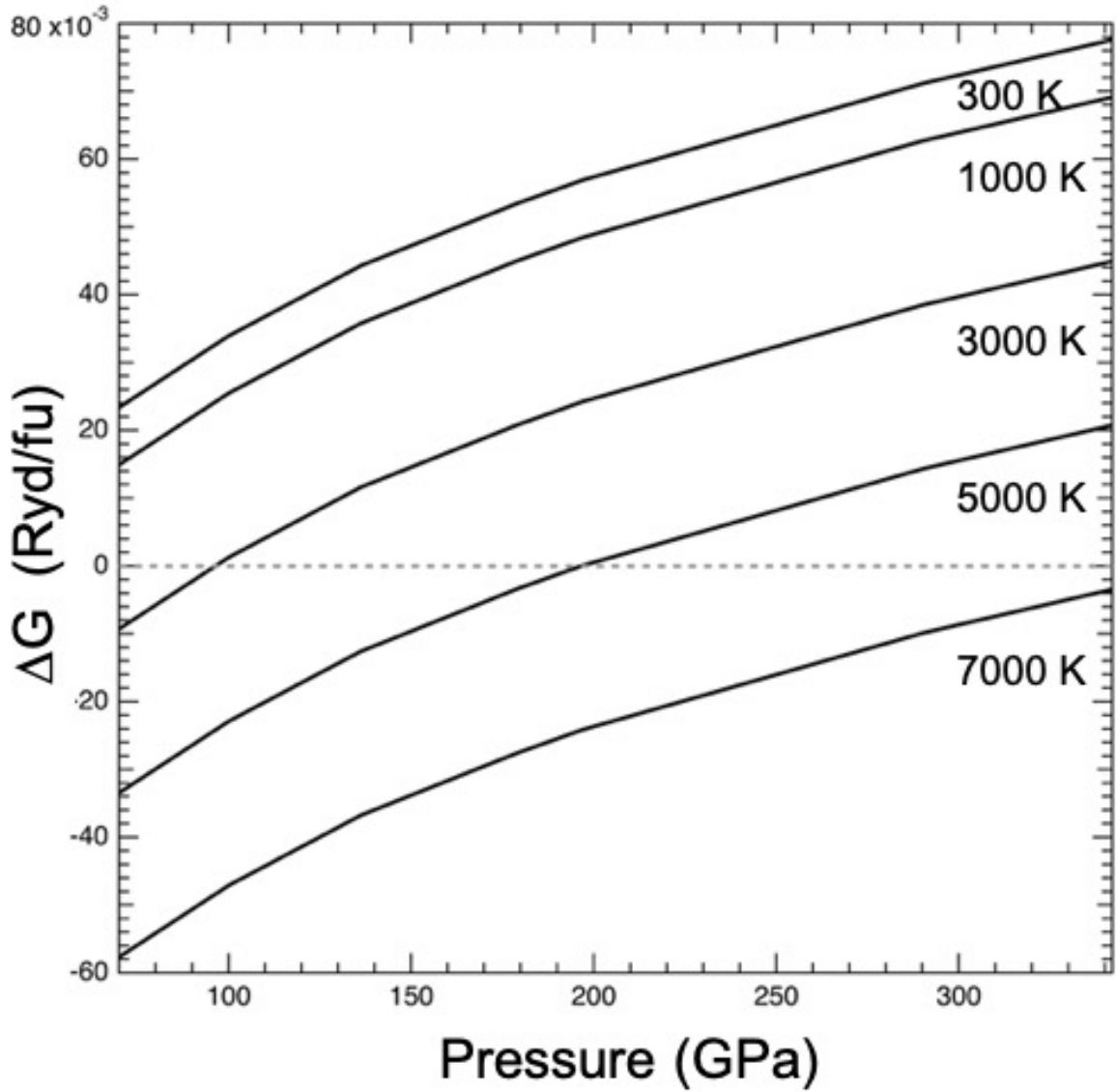

**Fig. S4** 224 atoms supercell for $I\bar{4}2d$-$Mg_2GeO_4$ before and after the phase transition at 200 GPa.

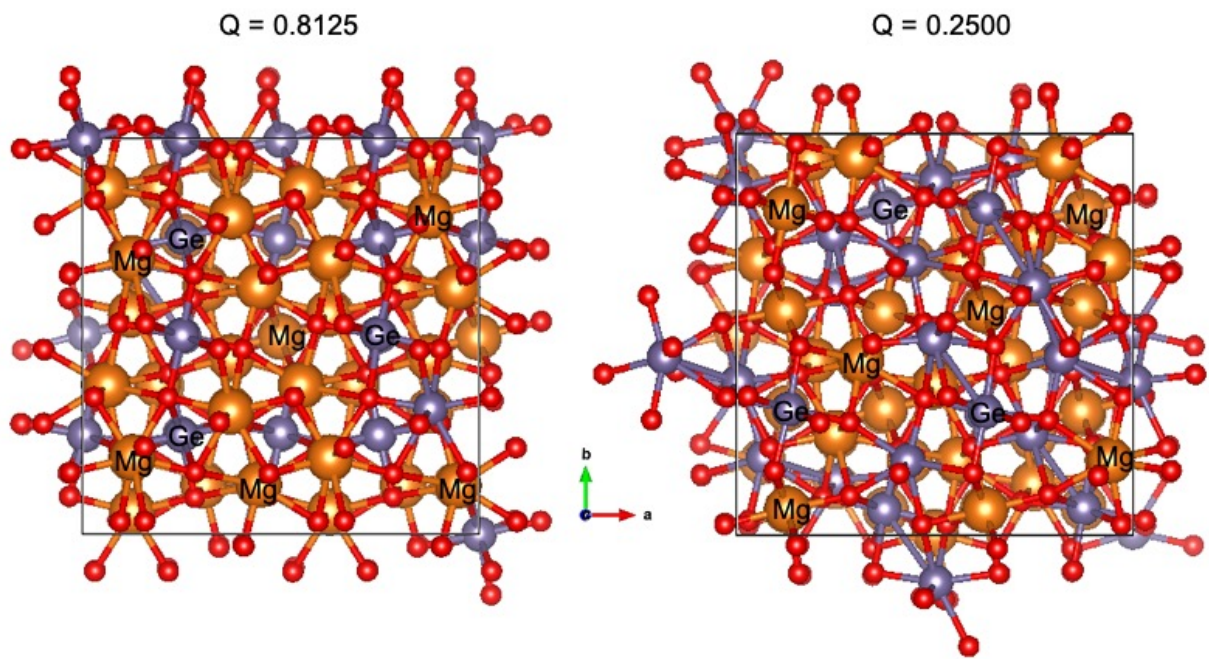

**Fig. S5** Enthalpy as a function of order parameter, Q for considered cluster sizes at 200 GPa. Table. S4 shows the $R^2$ values for the fits.

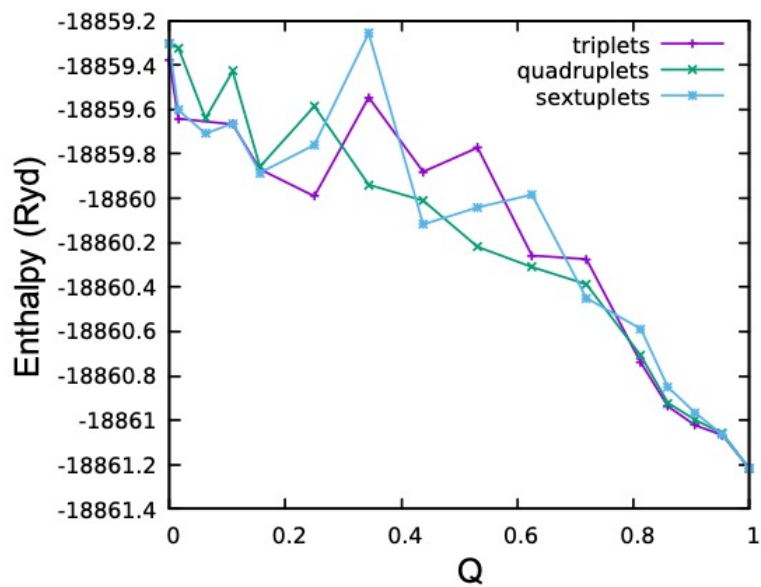

**Table S4.**

| Order of fit | $R^2$ adjusted | $R^2$ |
|---|---|---|
| 5 | 0.9371 | 0.944 |
| 4 | 0.9373 | 0.943 |
| 3 | 0.9277 | 0.932 |
| 2 | 0.925 | 0.928 |
| 1 | 0.882 | 0.885 |

**Fig. S6** Entropy versus order parameter at 200 GPa.

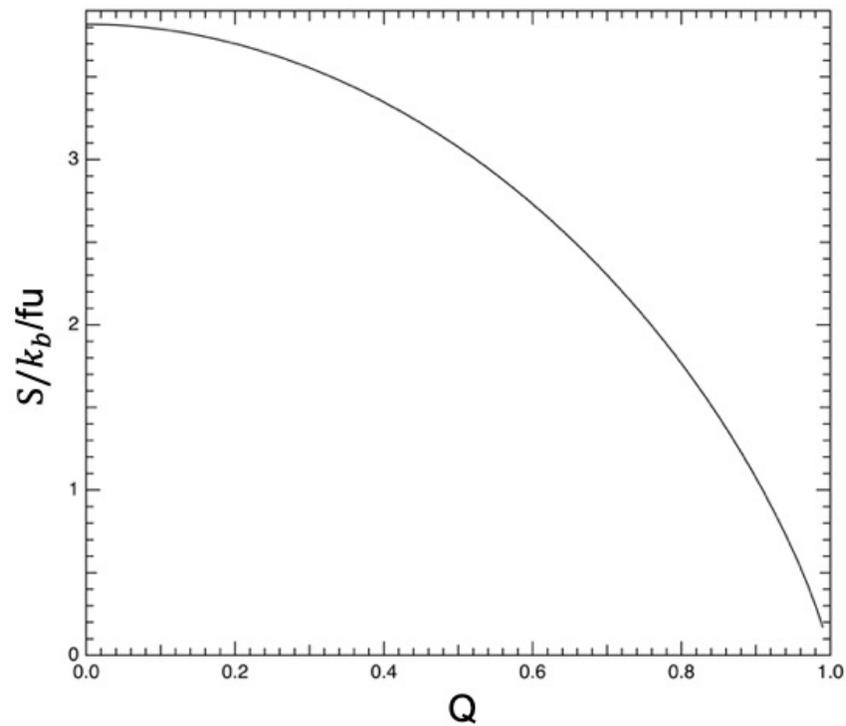

**Fig. S7** Change in Order parameter, Q with temperature at 200 GPa for 2nd (red), 3rd (blue) and 4th (green) order fits to the enthalpy vs Q.

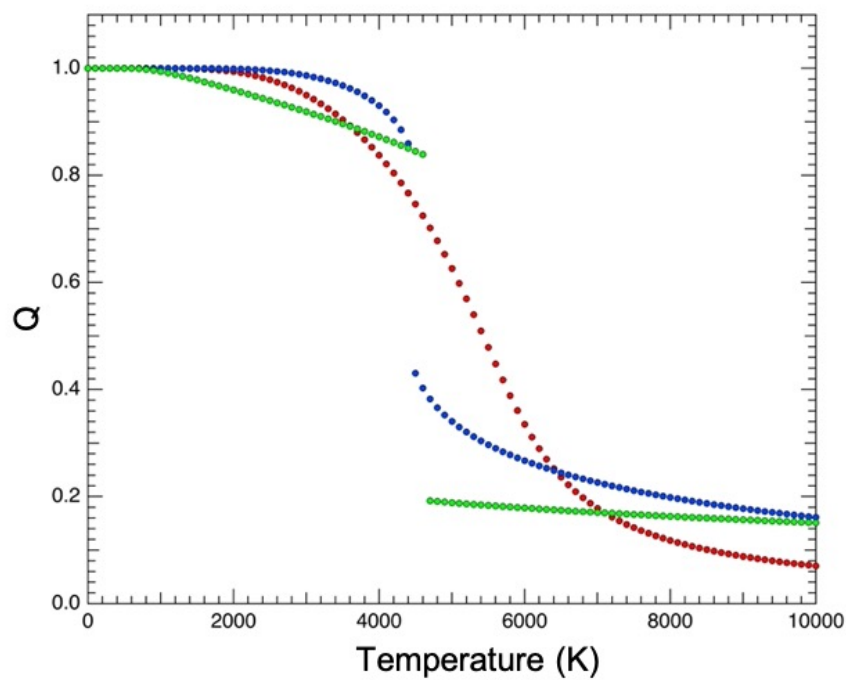

## S5. Supplemental X-Ray diffraction results for the Th$_3$P$_4$-type phase of Mg$_2$GeO$_4$

**Fig. S8.** X-ray diffraction patterns obtained on a sample heated at 158 GPa (H3) followed by subsequent cold compression and heating at 240 GPa (Run H3_2). The pattern at the bottom in grey shows the pattern obtained on quenching Mg$_2$GeO$_4$ after heating to 1806 K at 158 GPa showing weak peaks of the Th$_3$P$_4$ phase (indicated by red circles) formed metastably at this pressure. The upper traces (black) show a series of diffraction patterns at increasing temperature for the heating cycle at 240 GPa (H3_2). The peaks of the Th$_3$P$_4$ phase grow with temperature upon heating to as high as 3650 K, which supports the stability of the Th$_3$P$_4$ phase at these P-T conditions.

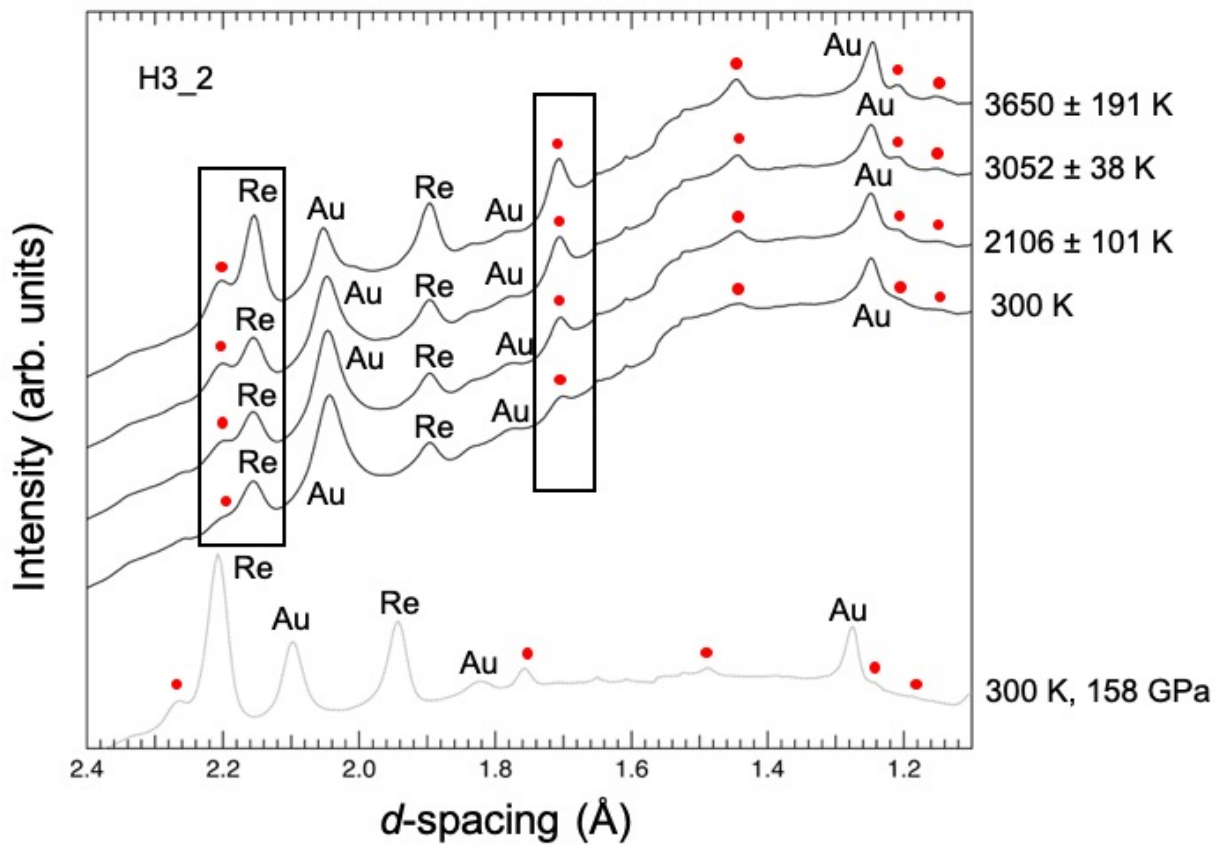

**Fig. S9.** Two-dimensional diffraction image of $Mg_2GeO_4$ after heating to 3650 K at 240 GPa with subsequent quenching to room temperature. The left figure shows the raw image, which is transformed to coordinates of azimuthal angle around the X-ray beam vs. two-theta in the right-hand image. The integrated one-dimensional pattern for this image is shown in Fig. 3. Arrows (yellow: Au, light blue: Re, and purple: $Th_3P_4$-type $Mg_2GeO_4$) show the peaks positions in the integrated pattern.

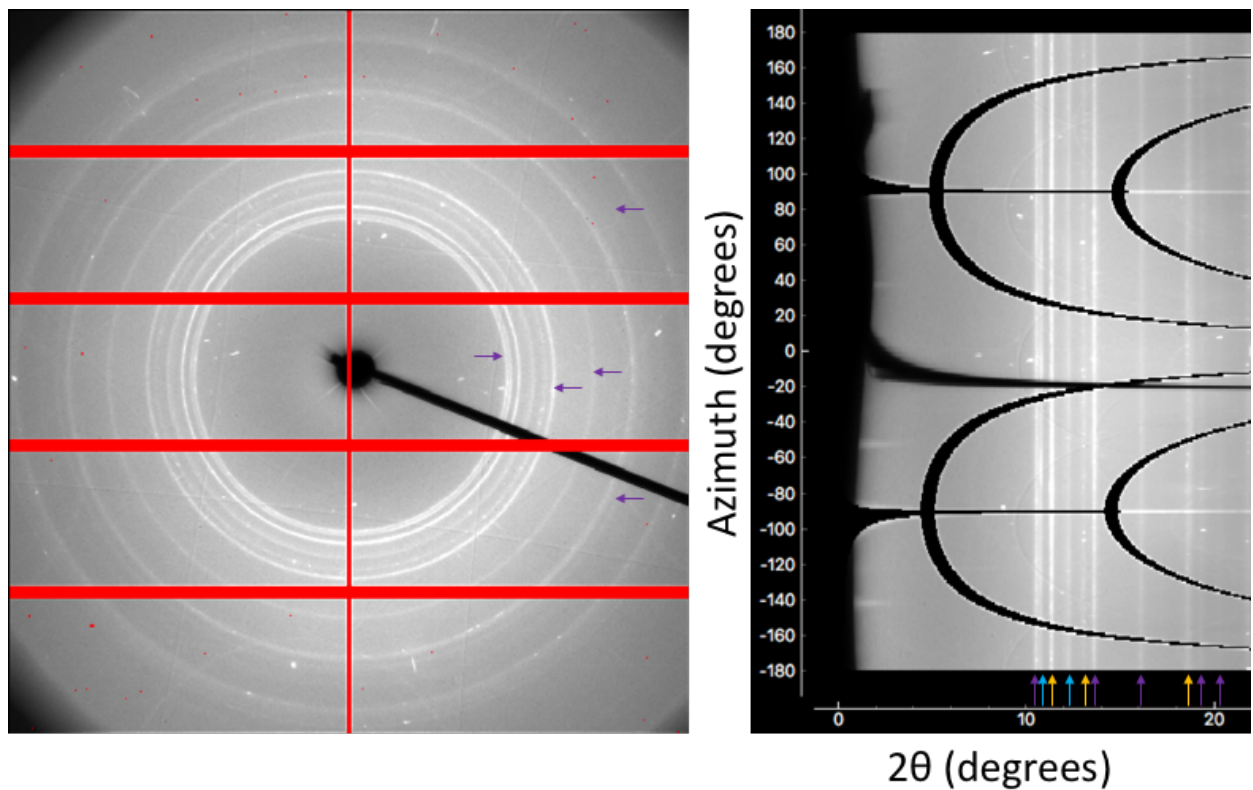

## S6. X-ray diffraction results for pPv MgGeO$_3$

**Fig. S10.** X-ray diffraction patterns during the heating cycle at 115 GPa (H1). Brown circles indicate post-perovskite peaks. The total heating duration was 11 minutes 21 seconds with ~1-1.5 minutes at each step.

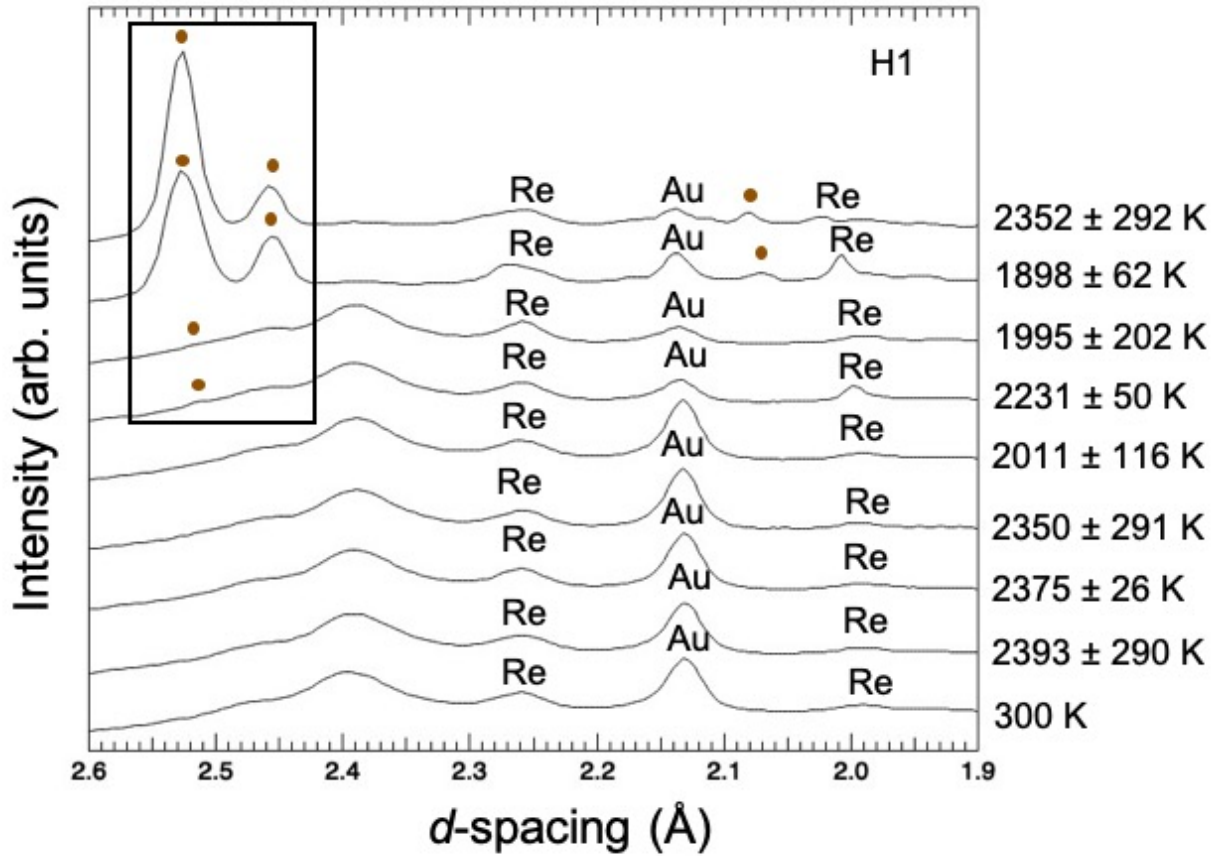

**Fig. S11.** X-ray diffraction pattern obtained after heating Mg$_2$GeO$_4$ to 3200 K at 179 GPa, followed by quenching to room temperature (quenched P = 172 GPa). The tick marks at the bottom indicate the expected peak positions for Th$_3$P$_4$ (red), post-perovskite (brown), gold (yellow), and MgO (blue). Miller indices of pPv-MgGeO$_3$ (brown) and Th$_3$P$_4$-type (red) Mg$_2$GeO$_4$ are listed above each peak.

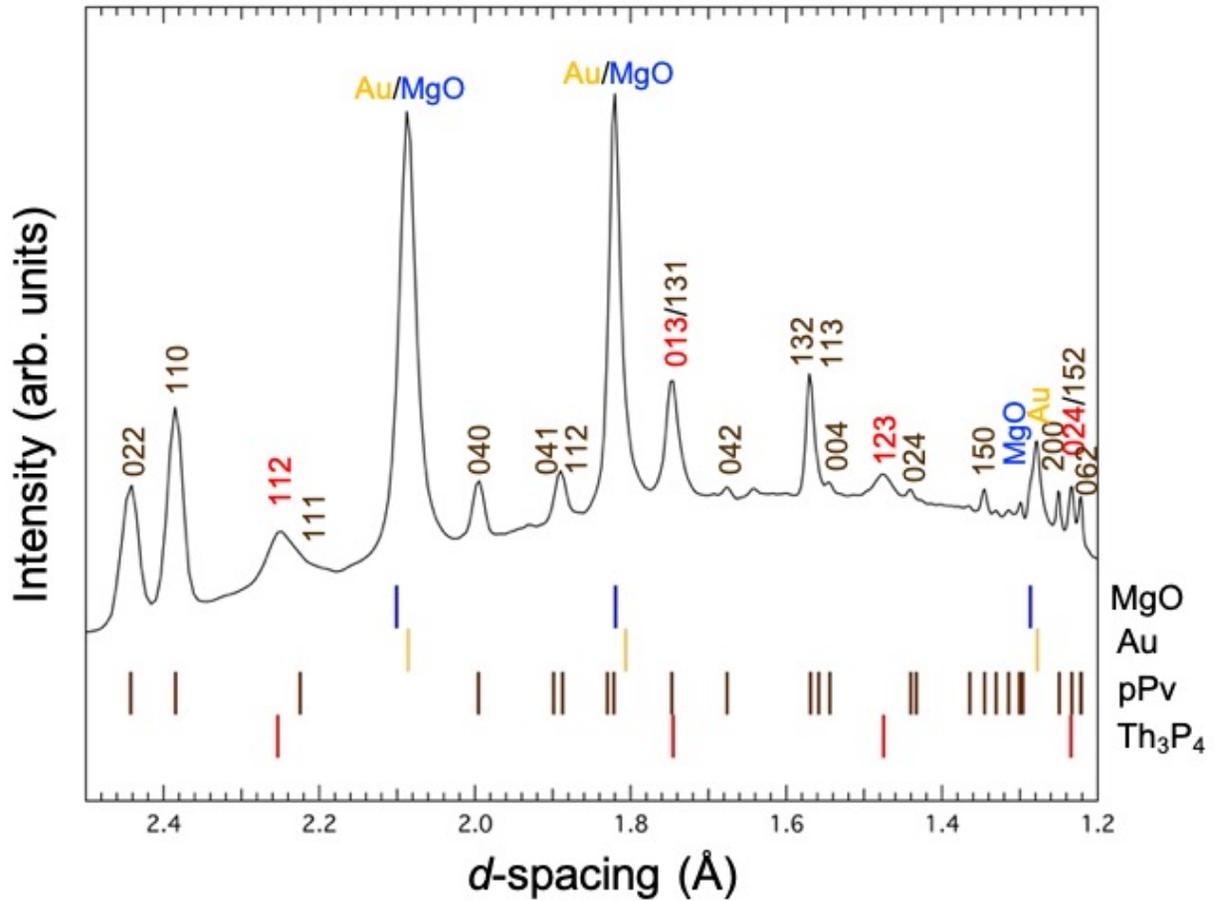

**Table S5.** Calculated and observed $d$-spacings of post-perovskite $Mg_2GeO_4$ at 172 GPa. The fit lattice parameters are $a$ = 2.498 (1) Å, $b$ = 7.982 (3) Å, and $c$ = 6.177 (4) Å.

| $h$ | $k$ | $l$ | $d$-observed (Å) | $d$-calculated (Å) | $\Delta d$ (Å) |
|---|---|---|---|---|---|
| 0 | 2 | 0 | 3.9905 | 3.9911 | -0.0006 |
| 0 | 0 | 2 | 3.0869 | 3.0883 | -0.0014 |
| 0 | 2 | 2 | 2.4431 | 2.4425 | 0.0006 |
| 1 | 1 | 0 | 2.3851 | 2.3847 | 0.0004 |
| 0 | 4 | 0 | 1.9956 | 1.9955 | 0.0001 |
| 1 | 3 | 1 | 1.7462 | 1.7471 | -0.0009 |
| 1 | 3 | 2 | 1.5694 | 1.5689 | 0.0005 |